\documentclass{llncs}

\usepackage{geometry}
\usepackage{geometry}
\usepackage{graphicx}
\usepackage{adjustbox}
\usepackage{multirow}
\usepackage{lipsum}
\usepackage{framed}
\usepackage{multirow}
\usepackage{lscape}
\usepackage{rotating}
\usepackage{tikz}
\usepackage{subfigure}
\usepackage{appendix}
\usepackage{todonotes}
\usepackage{alphalph}
\usepackage{manyfoot}
\usepackage{framed}
\usepackage{float}
\usepackage{longtable}
\usepackage{amsfonts}
\usepackage{todonotes}

\begin{document}
	
\title{A Value-driven Approach for Software Process Improvement - A Solution Proposal}

\author{Ramtin Jabbari
	\and Nauman bin Ali
	\and
	Kai Petersen}

\institute{Blekinge Institute of Technology, Karlskrona, Sweden\\
	\email{{rjb, nal, kps}@bth.se}\\
}

\maketitle              

\begin{abstract}   
Software process improvement (SPI) is a means to an end, not an end in itself (e.g., a goal is to achieve shorter time to market and not just compliance to a process standard). Therefore, SPI initiatives ought to be streamlined to meet the desired values for an organization. 

Through a literature review, seven secondary studies aggregating maturity models and assessment frameworks were identified. Furthermore, we identified six proposals for building a new maturity model. 

We analyzed the existing maturity models for (a) their purpose, structure, guidelines, and (b) the degree to which they explicitly consider values and benefits. Based on this analysis and utilizing the guidelines from the proposals to build maturity models, we have introduced an approach for developing a value-driven approach for SPI.

The proposal leveraged the benefits-dependency networks. We argue that our approach enables the following key benefits: (a) as a value-driven approach, it streamlines value-delivery and helps to avoid unnecessary process interventions, (b) as a knowledge-repository, it helps to codify lessons learned i.e. whether adopted practices lead to value realization, and (c) as an internal process maturity assessment tool, it tracks the progress of process realization, which is necessary to monitor progress towards the intended values.

\end{abstract}

\section{Introduction}
\label{sec:intro}
Increasingly the ability to deliver high-quality software is seen as a source for achieving and sustaining competitive advantage in the market. The role of software process to develop high quality software is well recognized \cite{softwareprocess}. Therefore, several proposals to design, evaluate and improve software development processes have been presented both from academia and industry. 

Several prescriptive and inductive approaches for software process assessment and improvement have been proposed \cite{GorscheksPItypes}. The prescriptive approaches like CMMI\footnote[1]{Capability Maturity Model Integration: A framework containing a process level improvement training and appraisal program that deals with the integration of maturity models in software development process, initiated by SEI, \textit{Software Engineering Institute} \cite{CMMIagile}.} and SPICE\footnote[2]{Software Process Improvement and Capability Determination: A set of standards for software development process to assess capability level of the processes in accordance with a reference model \cite{spice}. The SPICE standards present a descriptive model, based on which the process of activities would be assessed against fixed predefined levels of maturity \cite{spice2}.} tend to recommend a set of best practices identified from several organizations. The assessment frameworks for these approaches compare an organization's compliance with the recommendations in the standard. On the other hand, the inductive approaches like quality improvement paradigm (QIP) have a measurement focus. They start with understanding the current situation and then selecting, applying and evaluating improvements that are perceived as most useful for the specific organization.

The prescriptive approaches are criticized for their one-size-fits-all solution that does not take into account an organization's unique context. The inductive approaches by taking the challenges and the goals of an organization into account address this criticism. However, a problem in the inductive approaches is the lack of support they provide to practitioners in assessing their current process against the targeted values and in choosing which improvement actions to implement and in what order.

The main aim of this paper is to propose a value-driven approach to SPI that helps to overcome the limitations of inductive and deductive approaches. To achieve the aim, we built on existing work for developing maturity models with the following objectives:
\begin{enumerate}
\item Specifying the generic structure of maturity models presented in the literature in terms of their type, characteristics and specifications. From this we gain insights of how to structure a value-driven approach by understanding the limitations and strengths of the existing structures of maturity models.
\item Identifying the existing approaches, guidelines, and methods for designing a new maturity model proposed in the literature. Knowing the existing design approaches gives methodological guidance of how develop a new approach (in our case a value-driven approach to SPI).
\item Evaluating the existing approaches for achieving process maturity and the process assessment from the perspective of value and benefits. Understanding the evidence with respect to achieving value and benefits aids in identifying gaps as well as best practices that can be transferred to our solution.
\end{enumerate}

From a methodological point of view, we conducted a tertiary literature review to construct our solution. As a proof-of-concept we presented a benefits-dependency networks (BDN) for DevOps.  

The remainder of the paper is structured as follows: Section \ref{sec:background} describes the background of the study. Thereafter, we present the research questions and method in Section \ref{sec:researchmethod}. The results are presented in Section \ref{sec:researchresults}. Section \ref{sec:vadaspi} presents the results and our solution proposal (VASPI). Section \ref{sec:conclusionandfuture} concludes the paper and presents the directions for future research.

\section{Background and related work}\label{sec:background}
\subsection{Maturity Models and SPI}
In software engineering, maturity has been interpreted as the capability of implementing, developing, measuring, and improving software processes \cite{devmmslr,softwarematuritycapability}. Inspired by Crosby's quality management maturity grid \cite{crosby1980quality} several maturity models have been developed for organizations developing software \cite{maturitygrids,CMM}. 

Broadly, there are two different types of frameworks for software process assessment and improvement: inductive (descriptive) or prescriptive (comparative) \cite{devmmslr,GorscheksPItypes,typesofmm}. The first type, like quality improvement paradigm (QIP), focuses on a specific organization through realizing the current status and improving the identified challenges based on desired objectives. The second type (like CMMI) provide a set of predefined best practices based on successful experiences in other companies through making comparisons. 

CMMI, a typical example of prescriptive frameworks, has been widely used to assess the maturity and capability level of processes and organizations. It has two main model representations \cite{CMMI,CMMI2}: (1) a staged representation that uses levels to assess the overall maturity of an organization, and (2) a continuous representation that assesses the  capability levels of an organization in particular process areas. 

The staged model, sets the requirements for implementing a number of practices regardless of the nature and context of an organization. To address the criticism that not all practices are relevant for all companies, CMMI also suggests individual maturity levels for specific key process areas (KPAs) \cite{maturitygrids}. Therefore, using a continuous representation, a company can prioritize the improvement related to a particular area alone. This provides companies some flexibility compared to the staged model of organizational maturity. 

Pikkarainen and Mantyniemi \cite{plandrivenagile} have briefly explained how generic and specific goals and practices are positioned to measure processes: \textit{``Generic goals and practices apply to multiple process areas, whereas specific goals and practices apply to individual process areas. The specific goals and practices of process areas describe what kind of activities need to be carried out. Generic goals and practices are aimed at finding out how well the activities are performed''}.

The critical decision for a software organization is to select the best sequence of the `process areas' (risk management, causal analysis, and resolution) to fulfill the organizational needs and concerns. In addition to extra workloads, activities, documentation, and changes, an unsuitable set of the process areas can lead to negative results \cite{processareasorder}.

\subsubsection{Assessment Frameworks for Agile Software Development}
\label{subsubsec:agileapplicability}

Lots of practices have been introduced and used as part of agile software development methods, but the level of applicability of the practices and their desired benefits has not been properly systematized. Lack of standards or maturity models that specifically fit the agile practices could be a reason \cite{samireh}.

To evaluate the applicability of agile practices, few methods and guidelines have been introduced from different perspectives. McCaffery \textit{et al.} \cite{mccaffery2008ahaa} assesses the agility of development practices, particularly for small/ medium sized suppliers, by a combination of the flexibility of agile and traditional maturity models like CMMI and SPICE.

Further, Hossain \textit{et al.} \cite{hossain2009} have proposed a conceptual framework of strategies to select agile methods in the context of global software development (GSD). The framework was built based on challenges reported in the research literature on the use of agile methods in GSD.

Also, Soundararajan and Arthur \cite{soundararajan2011} have presented a framework to assess the adequacy, capability, and effectiveness of applying an agile method. The framework works through ``\textit{the identification of the agile objectives, principles that support the achievement of those objectives, and practices that reflect the spirit of those principles}'' \cite{soundararajan2011}. To this end, the relations between objectives and principles, and between principles and practices have been defined.

The \textit{Objectives, Principles} and \textit{Practices} (OPP) framework \cite{soundararajan2011} attempts to address the gap of other existing `agile assessment methods'. OPP framework assesses software and process artifacts by considering adequacy, capability, and effectiveness of an agile method through specifying the links between objective(s), principle(s), and corresponding practice(s).

\subsection{Value-Based Software Engineering}
Traditionally, software development has been done in a value-neutral manner \cite{vbse}. In reality, the decisions regarding software development require a trade-off between competing value attributes \cite{Barney2012}. The usefulness of pursuing a value-driven approach has been recognized and it has lead to several advancements in key knowledge areas of software engineering such as value-based requirements engineering and value-based product management \cite{khurumsoftwarevaluemap,vbse}. However, to the best of our knowledge no study has investigate the use of a value-based approach for SPI.

Khurum \textit{et al.} \cite{khurumsoftwarevaluemap} have presented an extensive catalogue of values for software development (referred to as software value map (SVM)). They have categorization the identified value aspects, sub-aspects, and value components. The catalogue provides a common understanding of values and presents the interrelationships between value perceptions from multiple perspectives  \cite{khurumsoftwarevaluemap}. 

In this study, we suggest the use of SVM \cite{khurumsoftwarevaluemap} as a means to identify values that are important to consider from an SPI perspective. These values will provide input to structure the knowledge related to process improvement using a Benefits Dependency Network (BDN) \cite{peppard2007managing}.

While the inductive approaches for SPI support the assessment and realization of the current state, but they do not help identifying improvement actions and their order. On the other hand, prescriptive approaches recommend one path to improve to pursue for all organizations \cite{devmmslr,typesofmm}. The proposed approach in this paper has the potential to address these weaknesses. By structuring the possible paths to the desired values using BDNs, any organization engaging in SPI can choose, track and pursue their own path for improvement actions. 

\subsection{Benefits Dependency Network}
\begin{figure}[!ht]
	\centering
	\includegraphics[width=\textwidth]{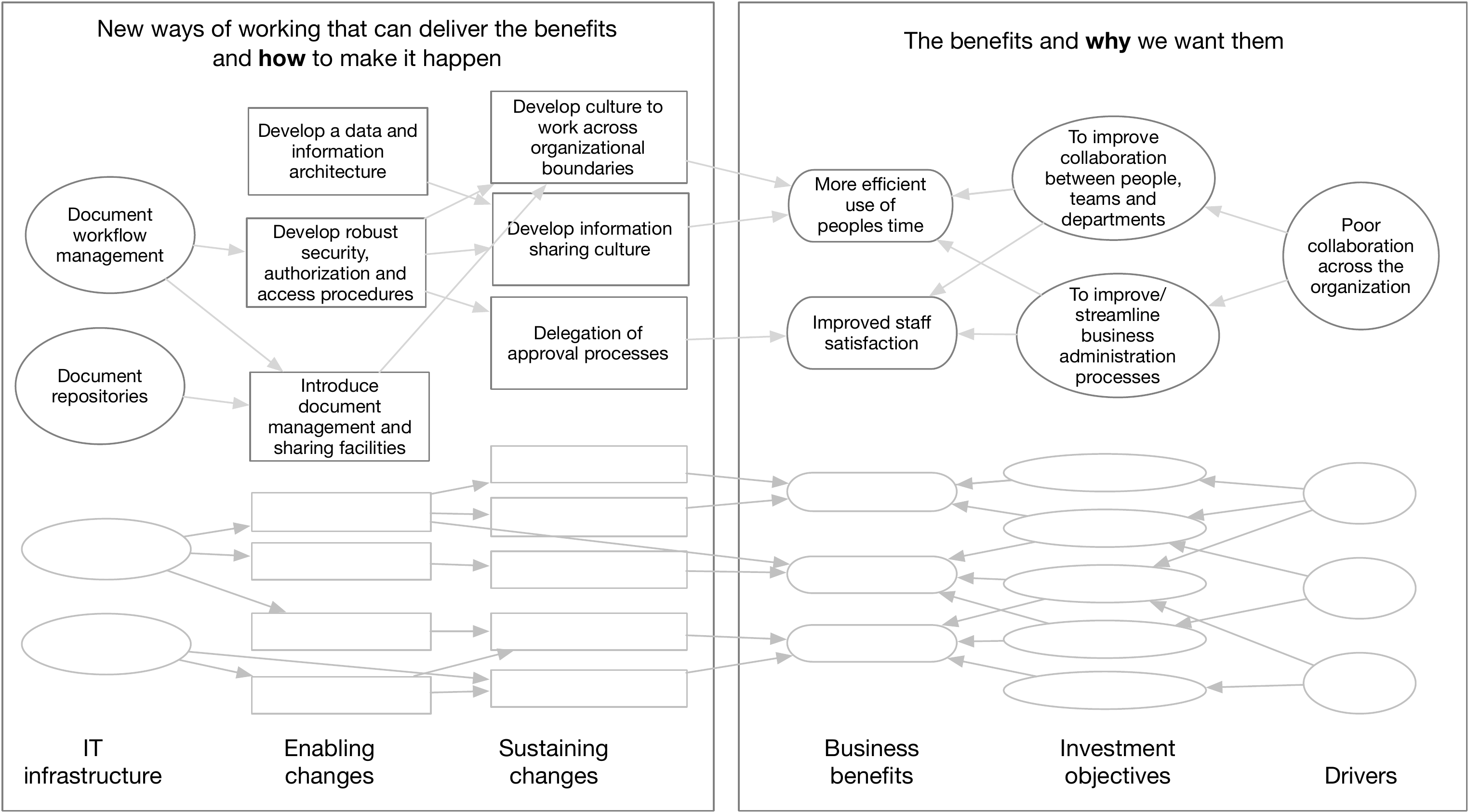}
	\caption{A template of a BDN from Peppard and Ward \cite{peppard2007managing}}
	\label{fig:BDNTemplate}
\end{figure}

Benefit dependency network (BDN) have been used to manage digital transformation in organizations. It has been used to identify, manage and deliver benefits from information technology investments \cite{peppard2007managing}. Figure \ref{fig:BDNTemplate} presents a template of a BDN.

We draw inspiration from BDNs to structure process related knowledge and streamline SPI initiatives. The visual map this approach provides gives the ability to develop a business case for SPI initiatives, and justifying what changes and improvements are required. The transparency will also help to understand, track and manage the actual progress implementing the improvements. 

\section{Research Method}
\label{sec:researchmethod}

In this section, we formulate the research questions and describe the method to address the research questions by the use of a literature review.

\subsection{Research Questions}
Four research questions (RQs) have been formulated for this study, as follows:

\textbf{RQ1. What is the generic structure of maturity models presented in the literature?}

RQ1 aims to identify the generic structure of maturity models, existing in the literature, to better understand the content and their relations to value/benefit-driven approaches. To address the RQ1, we have looked at existing studies, which explicitly discuss the topic of interest.

\textbf{RQ2. What are the approaches/guidelines/methods to design a new maturity model introduced in the literature?}

RQ2 aims to identify development processes presented in the literature to create a new maturity model. To address the RQ2, we looked at studies, which explicitly present a new or an enhancement of maturity model.

\textbf{RQ3. To what degree do existing maturity models take value/ benefit into account for assessment?}

The purpose of RQ3 is to investigate those maturity models, in which relevant values and benefits have been explicitly taken into account.
		
\textbf{RQ4. How could a new maturity model be proposed for a value-driven SPI approach?}

RQ4 aims to propose a new approach for creating a model to assess the maturity on the basis of values and benefits associated with practices in a specific context.
		
\subsection{Research Method}

To identify existing guidelines and procedures to develop a maturity model, we searched if secondary studies on the topic have already been conducted \cite{SLRguidelines}. We searched in Scopus and Google-Scholar using the following general search string:

\textit{(``capability model'' OR``maturity model” AND (``systematic review” OR ``research review” OR ``research synthesis” OR ``research integration” OR ``systematic overview” OR ``systematic research synthesis” OR ``integrative research review” OR ``integrative review” OR ``systematic literature review” OR ``systematic mapping” OR ``systematic map”)}

From this search, we identified seven secondary studies \cite{devmmslr,maturitymodelmappingstudy,10tarhan2016slr,lasrado2015maturity,von2010systematic,9spiSLR,13criticalreview} that are discussed in Section \ref{subsec:relatedreviews}. 

We re-analysed the relevant literature identified in these secondary studies to answer the research questions RQ1 -- RQ4.

\section{Results}
\label{sec:researchresults}

\subsection{Related Literature Review Studies}
\label{subsec:relatedreviews}

Table \ref{tab:relatedwork} presents the identified secondary studies we have used for extracting relevant information. 

\pagebreak
\begin{longtable}{ p{.16\textwidth} | p{.19\textwidth} | p{.106\textwidth} | p{.495\textwidth} | p{.05\textwidth} } 
	
	Focus/Goal	& Type of Study & \#Papers & Key Conclusion & Ref.\\ \hline
	
\small Identifying the state of the art of maturity models (The maturity of maturity models).
	
& \small Systematic mapping study \cite{SLRguidelines,Petersen2015guidelines}
	
& 237
	
& $\bullet$ \small \textbf{F1.} Maturity models mostly used in software development and engineering.

$\bullet$ \small \textbf{F2.} The most studies are about the development of maturity models proposing a new or an enhancement of existing ones.

$\bullet$ \small \textbf{F3.} Out of 128 studies related to the development of maturity models, 56 are conceptual and 42 are design-oriented.

$\bullet$ \small \textbf{F4.} The number of empirical or validated studies is scarce.

$\bullet$ \small \textbf{F5.} Those studies proposed new maturity models suffer from not reviewing existing models, not considering other available solutions, and not carefully examining their applicability and suitability in practice that negatively affect the quality and relevance of the new models.
	
& \cite{maturitymodelmappingstudy} \\ \hline
	
\small Classifying the existing maturity models.
	
& \small Systematic literature review \cite{SLRguidelines}
	
& 61
	
& $\bullet$ \small \textbf{F6.} 52 software process capability/maturity models are identified.

$\bullet$ \small \textbf{F7.} The identified models are classified in 29 specified domains like automotive systems, business process, etc.

$\bullet$ \small \textbf{F8.} 96\% of them (50 models) are developed based on CMMI and/or SPICE.
	
& \cite{von2010systematic} \\ \hline
	
\small Identifying the methods, practices, and activities used to propose or build maturity models.
	
& \small Systematic literature review
	
& 7
	
& $\bullet$ \small \textbf{F9.} Most of the new maturity models are based on CMMI, and SPICE.

$\bullet$ \small \textbf{F10.} The majority studies are from the system information domain.

$\bullet$ \small \textbf{F11.} Only one paper is about the development of maturity models in software engineering.

$\bullet$ \small \textbf{F12.} The majority of identified models are based on practical experiences, success factors, and practices in industry or company.

$\bullet$ \small \textbf{F13.} Lack of theoretical basis and distinct methodology.

$\bullet$ \small \textbf{F14.} Few proposed models are evaluated to assess the validity, reliability, and generalizability.

$\bullet$ \small \textbf{F15.} The general activities to build new maturity models are specified.
	
& \cite{devmmslr} \\ \hline
	
\small	Identifying the existing approaches used to develop maturity models and corresponding challenges during the development.
	
& \small Systematic literature review
	
& 34
	
& $\bullet$ \small \textbf{F16.} Five generic components identified to describe the structure of maturity models.

$\bullet$ \small \textbf{F17.} Recent studies spent significant effort to standardize the development of maturity models through \textit{prescriptive guidelines}, \textit{standardized vocabulary}, and \textit{validated procedure}.
	
$\bullet$ \small \textbf{F18.} Three main approaches identified as the development processes of maturity model (\textit{Meta models}).
	
$\bullet$ \small \textbf{F19.} New maturity models suffer from lack of empirical validation, operationalising maturity measurement, and theoretical foundations.
	
$\bullet$ \small \textbf{F20.} The challenges highlight potential gaps for further empirical research and validation on the topic.
	
& \cite{lasrado2015maturity} \\ \hline
	 
\small Investigating the current software process improvement (SPI) approaches using in small and medium firms (SMEs)
	
& Systematic literature review
	
& 11
	
& $\bullet$ \small \textbf{F21.} Most SMEs suffer from a lack of knowledge and resources to adopt proper SPIs.

$\bullet$ \small \textbf{F22.} Although the majority of SPIs for SMEs are on the basis of CMMI, these standards are developed primarily for larger firms that do not fit the SMEs \cite{9spiSLR,29spiforvse}.

$\bullet$ \small \textbf{F23.} Adopting the mentioned standards in SMEs leads to costly time and resource consumptions.

$\bullet$ \small \textbf{F24.} Lack of generalizability for those SPIs proposed for a specific SME.

$\bullet$ \small \textbf{F25.} There is a vital need for the development of SPIs in particular for SMEs.
	
& \cite{9spiSLR} \\ \hline 
	
\small Studying the strengths and limitations of existing maturity models
	
& Literature review
	
& \textit{not specified}
	
& $\bullet$ \small \textbf{F26.} There is no consistent definition and consensual perception of maturity models in the literature.

$\bullet$ \small \textbf{F27.} Crosby’s Quality Management Maturity Grid is one of the first staged maturity models for quality assessment that could recognize the importance of human factors like leadership and collaboration.

$\bullet$ \small \textbf{F28.} Bessant’s Continuous Improvement Capability model, as another staged model, is able to assess the general level of maturity and indicate how to conduct continuous improvement for enhancing the performance by providing a roadmap.

$\bullet$ \small \textbf{F29.} CMMI is the most used staged model, share some similarities in terms of structure and content.

$\bullet$ \small \textbf{F30.} Maturity models which are ``\textit{step-by-step recipes}'' suffer from lack of theoretical foundation, empirical basis and reality, validity, and generalization such as Crosby's Grid, Bessant, and CMMI.

$\bullet$ \small \textbf{F31.} There is a large number of similar models based on CMMI without satisfactory documentation, thoughtful adoption, and economic basis.

$\bullet$ \small \textbf{F32.} A shortage of guidelines for specific steps and activities towards the improvement of maturity levels (Bessant's Model and CMMI).

$\bullet$ \small \textbf{F33.} The majority of the models are costly and complicated to be applied and they are more suitable for large and bureaucratic organization (CMMI).

$\bullet$ \small \textbf{F34.} CMMI mostly focuses on process and disregards the importance of people, culture, and organization.
	
& \cite{13criticalreview} \\ \hline
	
\small Understanding the status of Business Process Maturity Models (BPMM)
	
& Systematic literature review
	
& 61
	
& $\bullet$ \small \textbf{F35.} 20 BPMM identified, out of which 9 models are considered as the leading ones in terms of the rate of attention in the literature.

$\bullet$ \small \textbf{F36.} Only 2 out of 9 models have been empirically studied and validated.

$\bullet$ \small \textbf{F37.} Only 7 studies out of 61 reported the validation.

$\bullet$ \small \textbf{F38.} A lack of validation negatively influences the acceptance level of the models by practitioners and potential audience.

$\bullet$ \small \textbf{F39.} Only two leading models distinguish between \textit{maturity model} (representing a path for improvement) and \textit{assessment model} (specifying the current situation) that indicates a lack of consistency in terms of design, scope, and terminology.

$\bullet$ \small \textbf{F40.} A lack of theoretical and empirical studies on existing maturity models.
	
& \cite{10tarhan2016slr} \\ \hline

\caption{Related Systematic Studies} 
\label{tab:relatedwork}
\end{longtable}

\subsection{Values/Benefits versus Maturity}
\label{subsec:valuebenefitmaturity}

\subsubsection{The Characteristics of Quality for Software Engineering Processes:}
\label{subsubsec:characteristicsofqualityfor}

Kroeger \textit{et al.} \cite{22valuematurity} conducted a study by interviewing 17 experienced software engineers in addition to reviewing 30 peer-reviewed studies from SPI case studies. They aimed to identify and characterize the quality criteria of software engineering processes. 

To this end, they have specified three main quality evaluation methods: (1) Product and process measurement, (2) Software process assessment, and (3) Process modeling and analysis.

The first category contains methods which indirectly assess the quality of processes by measuring the developed systems and their outputs. The second category is an evidence-based approach that compares the processes with a set of standard best practices (like CMMI). The third determines the quality of processes by the use of modeling and analysis such as \textit{petri-net based formalism, visual process language}, and \textit{unified modeling language} that deals with the processes in terms of a formal language. Then, through analysis techniques, inconsistencies, ambiguities, and potential issues would be identified before the process is deployed.

In fact, different approaches focus on different aspects of specific quality attributes. For example, the categories (1) and (3) mostly focus on \textit{efficiency} and \textit{effectiveness}, while the category (2) takes \textit{suitability} and \textit{manageability} into consideration.

To characterize the quality of software processes, Kroeger \textit{et al.} \cite{22valuematurity} have listed the most frequently discussed quality attributes. In total, 15 attributes extracted from both interviews and literature: Effectiveness 88\% \footnote{The percentages are only related to the interviews. Though, \textit{effectiveness} was also the most frequently discussed one in the literature \cite{22valuematurity}.}, Adaptability 76\%, Compatibility 59\%, Applicability 53\%, Deployability 47\%, Efficiency 47\%, Monitorability 41\%, Understandability 41\%, Controllability 35\%, Learnability 29\%, Predictability 29\%, Analyzability 29\%, Accessibility 24\%, Modifiability 24\%, and Supportability 24\%.

\subsection{Methods/Guidelines to Design Maturity Models}
\label{subsec:otherguidelines}
In this section, we attempt to identify studies that propose new maturity models, present the steps taken to design them, and discuss the motivations or problems to be addressed. Table \ref{tab:methodsguidelinesMM} shows the identified literature used for this study. 

\begin{table}[!ht]
\centering
\caption{The Studies Proposing Methods/Guidelines to Design Maturity Models}
\label{tab:methodsguidelinesMM}
\scalebox{1}{
\begin{tabular}{p{12cm} c c c}

\textbf{Goal} & \textbf{Steps} & \textbf{Motivation} & \textbf{Ref.} \\ \hline
 
Developing a domain-specific maturity model & \checkmark & \checkmark & \cite{27focusareaMM} \\ \hline

Building SPI in very small enterprises & \checkmark & \checkmark & \cite{29spiforvse} \\ \hline

Measuring the maturity of requirements engineering process & \checkmark & \checkmark & \cite{niazi2005} \\ \hline

Developing a maturity model for SPI implementation through adapting CMMI and Beecham's model.\footnotemark[1] & \checkmark & \checkmark & \cite{niaziSPImaturity} \\ \hline

Proposing a maturity model for the assessment and improvement of software maintenance activities. & \checkmark & \checkmark & \cite{aprilsmmm} \\ \hline

Developing a roadmap for CollabMM\footnotemark[2] evolution through the lesson learned of model applications and literature review. & \checkmark & \checkmark & \cite{24CollabMM} \\ \hline

\end{tabular}}
\end{table}

\footnotetext[1]{A requirements process improvement model presented by Beecham \textit{et al.} \cite{beecham}.}
\footnotetext[2]{Collaboration Maturity Model (CollabMM) \cite{24CollabMM}.}

\section{Discussions and Construction of VASPI}
\label{sec:vadaspi}

This section discusses the research questions based on the results. We also present our solution proposal for a value-driven approach for software process improvement (VASPI).

\subsection{RQ1. \textit{What is the generic structure of maturity models presented in the literature?}}

First, we reviewed the structure of maturity models 
presented in the literature through identifying 
type, characteristics, and specification (levels of maturity, descriptors, dimensions, and activities). The majority of discussions on maturity models have been concentrated on the type \textit{staged models} like Crosby, Bessant, and CMMI \cite{maturityclassification,crosby,maturitygrids,CMM,lasrado2015maturity,13criticalreview}.

In fact, as a key conclusion of the reviewed papers (F8, F9, F22, and F31 in Table \ref{tab:relatedwork}), the majority of identified models are based on CMMI, and SPICE. The 
common characteristics identified are as follows:
\begin{itemize}
\item Number of levels. (For example, CMMI has five levels of maturity.)
\item Descriptor for each level. (CMMI has five different descriptors: Initial, Managed, Defined, Quantitatively managed, and Optimizing.)
\item Description of each descriptor.
\item Number of dimensions (For example, the process areas in CMMI).
\item Number of elements or activities for each dimension.
\item Description of element or activity performed at each maturity level.
\end{itemize}

Then, we looked at the structure of CMMI, as the most used staged maturity model (See section \ref{subsubsec:characteristicsofqualityfor}). The existence of boundaries between CMMI process areas makes it hard to fit, e.g., agile practices in which it is not possible to separate the areas explicitly. The process areas also could not support overlaps between activities and dependencies among the areas.

In fact, the staged models with fixed predefined levels of maturity suffer from a lack of ability for expressing dependencies among processes within a capability level, while flexible models could provide, e.g., more than five maturity levels. Hence, flexible models allow the definition of intermediate status that indicates further details of goals, practices, potential improvements, and maturity \cite{devmmslr,27focusareaMM,29spiforvse}. Although, it brings a new challenge for organizations to use CMMI (as an indicator for software development process maturity) and agile practices to take benefits at the same time \cite{mccaffery2008ahaa}.

As a consequence, a tendency exists towards customizing maturity models or proposing new models in accordance with specific domains, which means there is no one model that fits all \cite{von2010systematic}.

\subsection{RQ2. \textit{What are the approaches/guidelines/methods to design a new maturity model introduced in the literature?}}

From the literature, we found studies which propose new maturity models most often suffer from avoiding existing models and solutions, a lack of theoretical foundation, empirical evaluation, and validity (F5, F13, F14, F19, F20, F36, F37, F38 and F40 in Table \ref{tab:relatedwork}).
Therefore, before proposing our solution, we attempted to identify the methods presented for building a new maturity model by the literature. Although the identified studies do not present detailed steps of the development, or they belong to the specific context that is not possible to be generalized, we could specify key steps of the development procedure. We aggregated generic activities performed to create a new (or an enhancement of) maturity model. To this end, we identified six studies compared and shown in Table \ref{tab:comparisonsteps}. For the comparison, six characteristics have been defined as follows:
\begin{itemize}
\item \textbf{Context.} A \textit{context} for which a model has been specifically proposed.
\item \textbf{Origination.} The basis on which a model has been developed. We have identified four originations used in the literature to build maturity models: (1) Practical experience (the most available models are developed based on practical experience \cite{13criticalreview}). (2) Standards like CMMI and SPICE. (3) Knowledge engineering. (4) Design science.
\item \textbf{Source.} It specifies an existing model, framework, or structure based on which a model has been proposed.
\item \textbf{Type.} Two main types of maturity model distinguished in the literature: (1) Staged (fixed models) which have predefined and distinct levels of maturity (2) Flexible models which can recognize interdependencies among different levels of maturity.
\item \textbf{Meta-model \cite{lasrado2015maturity}.} Three main approaches identified as procedures for maturity model development: a \textit{6-phase}, \textit{8-step}, and \textit{5-step} development process based on the concept of maturity models \cite{lasrado2015maturity}.
\item \textbf{Evaluation.} We considered whether or not the proposed model has been evaluated and validated.
\end{itemize}

\begin{table}[!ht]
\centering
\caption{Comparison between studies of proposing a maturity model}
\label{tab:comparisonsteps}
\scalebox{1}{
\begin{tabular}{c c p{2.51cm}p{3.1cm}p{2.6cm} c c c}

\textbf{\#} & \textbf{Study} & \textbf{Context} & \textbf{Origination}\cite{devmmslr} & \textbf{Source} & \textbf{Type} & \textbf{Meta-model} & \textbf{Evaluation}\\ \hline

1 & \textbf{\cite{27focusareaMM}} & Focus Area & Practical experience \& Design science & \textit{not reported} & Flexible & I\footnotemark[1] & Yes\\ \hline

2 & \textbf{\cite{29spiforvse}} & Very small enterprises & Standards & CMMI, SPICE & Staged & $\bullet$ & Yes\\ \hline

3 & \textbf{\cite{niazi2005}} &Requirements Engineering& Practical experience & RMM\footnotemark[2] & \textit{not reported} & $\bullet$ & Yes\\ \hline

4 &\textbf{\cite{niaziSPImaturity}} &SPI Implementation& Practical experience \& Standards & CMMI, RPI\footnotemark[3] & Staged & $\bullet$ & Future planned\\ \hline

5 & \textbf{\cite{aprilsmmm}} & Maintenance & Practical experience \& Standards & CMMI & Staged & $\bullet$ & Yes\\ \hline

6 & \textbf{\cite{24CollabMM}} & Collaboration & Practical experience \& Standards & CMMI, BPMM\footnotemark[4], KMMM\footnotemark[5] & Staged & I & Yes\\ \hline

\end{tabular}}
\end{table}
\footnotetext[1]{Meta-Model I \cite{lasrado2015maturity}}
\footnotetext[2]{Requirements Maturity Model presented by Sommerville \textit{et al.} \cite{sommerville}}
\footnotetext[3]{Requirements Process Improvement model presented by Beecham \textit{et al.} \cite{beecham}}
\footnotetext[4]{Business Process Maturity Model}
\footnotetext[5]{Knowledge Management Maturity Model}

To specify a generic procedure for developing maturity models, we have aggregated the findings from the literature \cite{29spiforvse,27focusareaMM,niazi2005,niaziSPImaturity,aprilsmmm,24CollabMM}. We considered the most common activities and corresponding methods to build maturity models. Consequently, four main phases have been defined:

\begin{itemize}
\item[(1)] \textbf{Motivation:} Through evaluating specific contexts and/or existing models, the motivation has been gained to build a new model.
    \begin{itemize}
    \item Evaluate the \textit{context}, for which a model would be specifically proposed through different evaluation methods like survey, literature review, etc.
    \item Evaluate existing model(s).
    \end{itemize}
\item[(2)] \textbf{Definition:}
    \begin{itemize}
    \item Identify gaps.
    \item Address the gaps.
    \end{itemize}
\item[(3)] \textbf{Construction:}
    \begin{itemize}
    \item Propose new model.
    \item Present an enhancement of existing model(s).
    \end{itemize}
\item[(4)] \textbf{Evaluation:}
    \begin{itemize}
    \item Evaluation/Validation.
    \end{itemize}
\end{itemize}

\begin{figure}[!ht]
\centering
\includegraphics[width=0.9\textwidth]{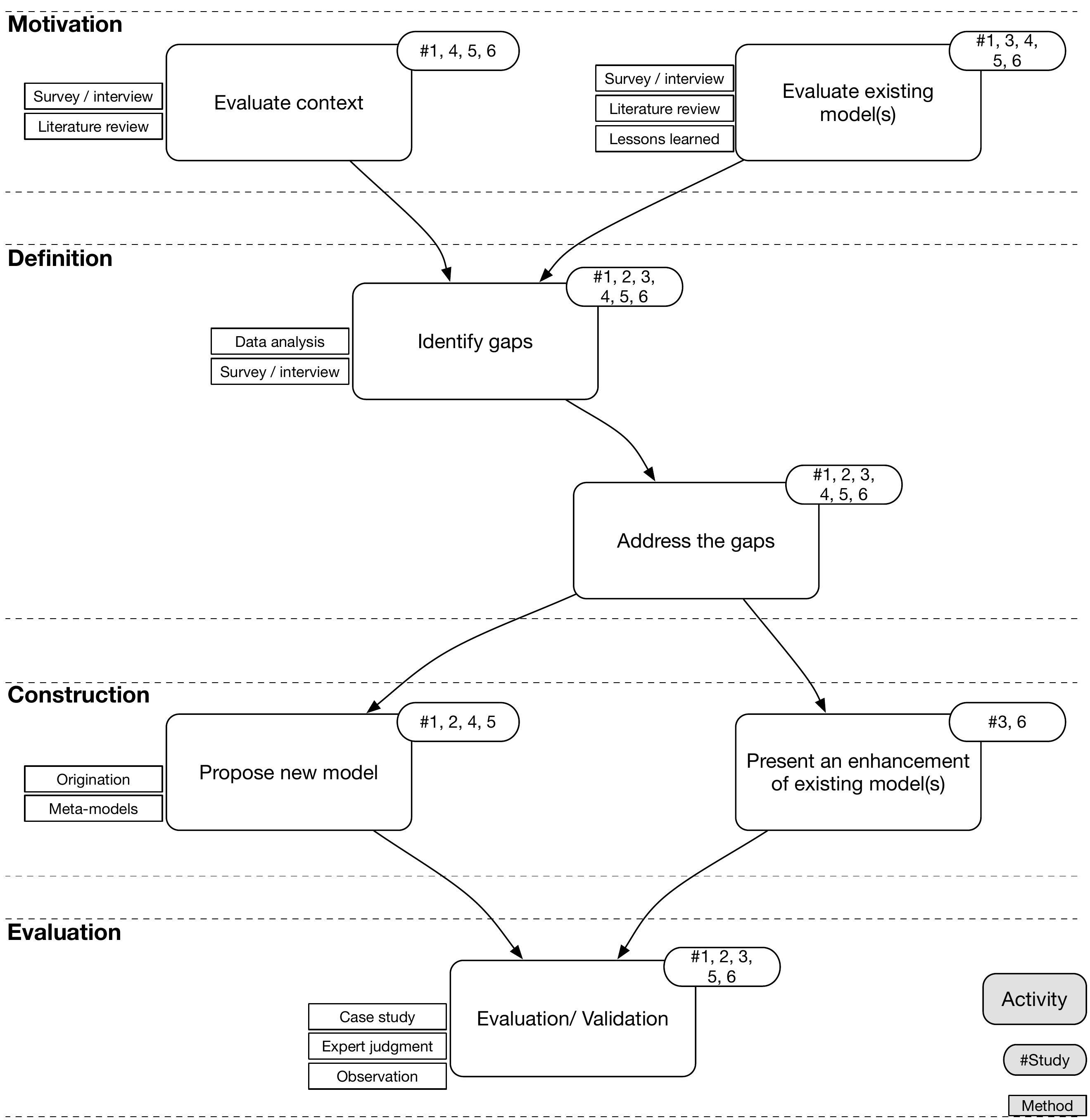}
\caption{Common procedure of creating maturity models}
\label{fig:maturitymodelprocedure}
\end{figure}

Figure \ref{fig:maturitymodelprocedure} indicates the generic development procedure for maturity models identified from the studies. Numbers of the activities show the references presented in Table \ref{tab:comparisonsteps}.

\subsection{RQ3. \textit{To what degree do existing maturity models take value/ benefit into account for assessment?}}

To our knowledge, the number of studies about maturity and actual benefits is scarce. Most existing discussions interpret the topic as general quality aspects and attributes of the software. Kroeger \textit{et al.} \cite{22valuematurity} stated that the majority of existing methods used to assess quality attributes of software process development have been limited to indirect measurement and evidence-based models (CMMI and formal models). These methods do not consider actual relations between practices and corresponding values, benefits, and quality attributes.

Moreover, such approaches mostly focus on general software quality attributes like \textit{efficiency}, \textit{effectiveness}, \textit{suitability}, and \textit{manageability}, which are related to the whole software development process and not associated to every single practice within the process.

Also, some maturity models like CollabMM proposed by Magdaleno \textit{et al.} \cite{24CollabMM} consider quality attributes and benefits as advantages of the model usage. For example, it has been claimed that by applying CollabMM, time reduction, less complexity, learning, satisfaction, and innovation would be achieved. But, CollabMM does not specify the relations between practices and their benefits.

Some challenges also realized in relation to agile approach which are: A lack of measures or assessment tools that practically fit agile practices \cite{samireh} and a lack of maturity models that consider the relations (dependencies) among agile practices (for example, iteration review requires iterative development to be in place \cite{iterativedev}) and between the practices and values/benefits.

As mentioned in section \ref{subsubsec:agileapplicability}, the OPP framework (Objectives, Principles, and Practices) considers the relation between objectives, principles, and practices. However, it does not take the relations/dependencies among practices into account \cite{soundararajan2011}.

\begin{figure}[!ht]
\centering
\includegraphics[width=0.8\textwidth]{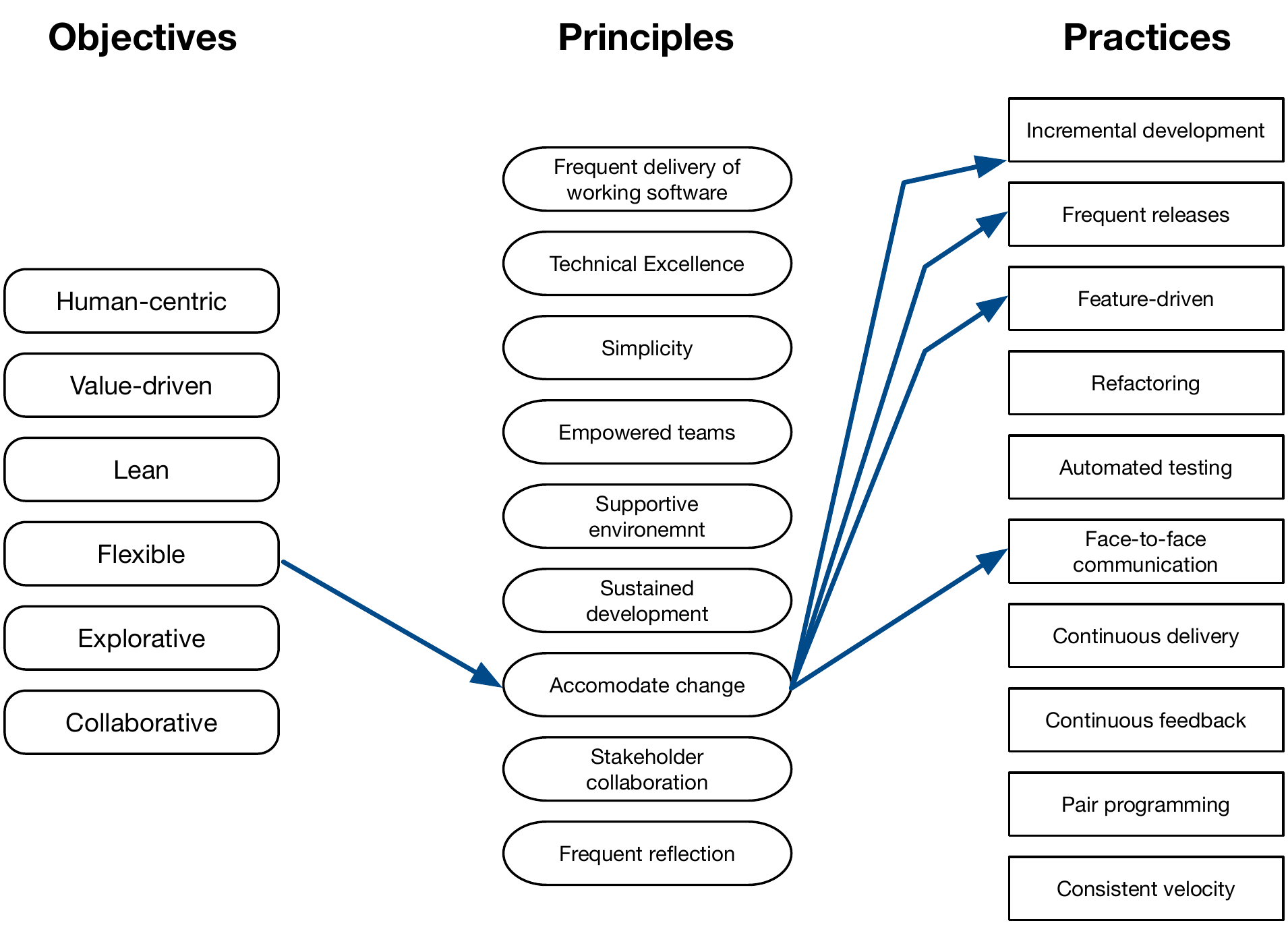}
\caption{An example of OPP framework}
\label{fig:oppframework}
\end{figure}

Figure \ref{fig:oppframework} shows an example of OPP framework that indicates accommodate change (as an agile principle) is linked to a number of practices. (Incremental development, Frequent releases, Feature-driven, and Face-to-face communication). As Figure \ref{fig:oppframework} shows, relations among a group of practices are missing (i.e., practices dependencies or a sequential order).

Moreover, the level of objectives and principles are too high that does not consider actual values/benefits which are explicitly coming from a specific practice. Hence, making decisions regarding applying particular practices becomes hard and negatively affects the feasibility of the framework.

Therefore, there is a lack of model/framework to assess the whole development process containing a set of practices following the desired benefits. Also, the majority of existing models are fixed model type and have rigid levels of maturity. Consequently, they are ineffective to improve processes practically, as they cannot express interdependencies among the processes for improving a maturity level \cite{25baars2016,27focusareaMM}. On top of these challenges, these models are too expensive and complicated to be implemented \cite{29spiforvse,9spiSLR}.

To address the identified gaps of existing maturity models, we realized the need for a new concept of maturity that meets the lack of actual relations between values/benefits and practices and has a consolidated view over the whole process of software development (See section \ref{subsubsec:newmaturitymodel}). 

\subsection{RQ4. \textit{How could a new maturity model be proposed for a value-based assessment approach?}}
This section presents our proposed model/framework as a guideline to design a new maturity model for software engineering development.
\subsubsection{Purpose:}
\label{subsubsec:purpose}
We aimed to propose a set of guidelines to design a model/framework for maturity assessment of applied software development practices based on our literature review. This guideline considers the relations among practices (i.e., dependencies among them) and corresponding goals/values/benefits.
\subsubsection{Need of a new approach for maturity model:}
\label{subsubsec:newmaturitymodel}
To address the identified gaps come from existing maturity models, the proposed guideline contributes through creating a value-driven model, as follows:
\begin{itemize}
\itemsep0em
\item To meet a lack of actual relations between goals/values/benefits and practices.

Instead of general appraisal characteristics, it considers the relations between specific practices and desired benefits.

\item To provide a holistic perspective of practices that includes relations/dependencies among the practices and the link to the desired goals/values/benefits. 

\item To provide a maturity model/framework to do only what is needed to be done to achieve the desired goal(s)/value(s)/benefit(s).

Instead of applying general processes of assessment, the proposed guideline in this study aims to design a new maturity model/framework based on the actual practices that need to be conducted in accordance with the desired benefits.
\end{itemize}
\subsubsection{Development:}
\label{subsubsec:development}

Two main approaches are presented by our guideline: Literature approach (L) and in-Practice approach (P).

Figure \ref{fig:guidelineoverview} shows the guideline proposed in this study to design a new maturity model. We describe the process of developing the guideline, as follows:
\begin{figure}[!ht]
\centering
\includegraphics[width=\textwidth]{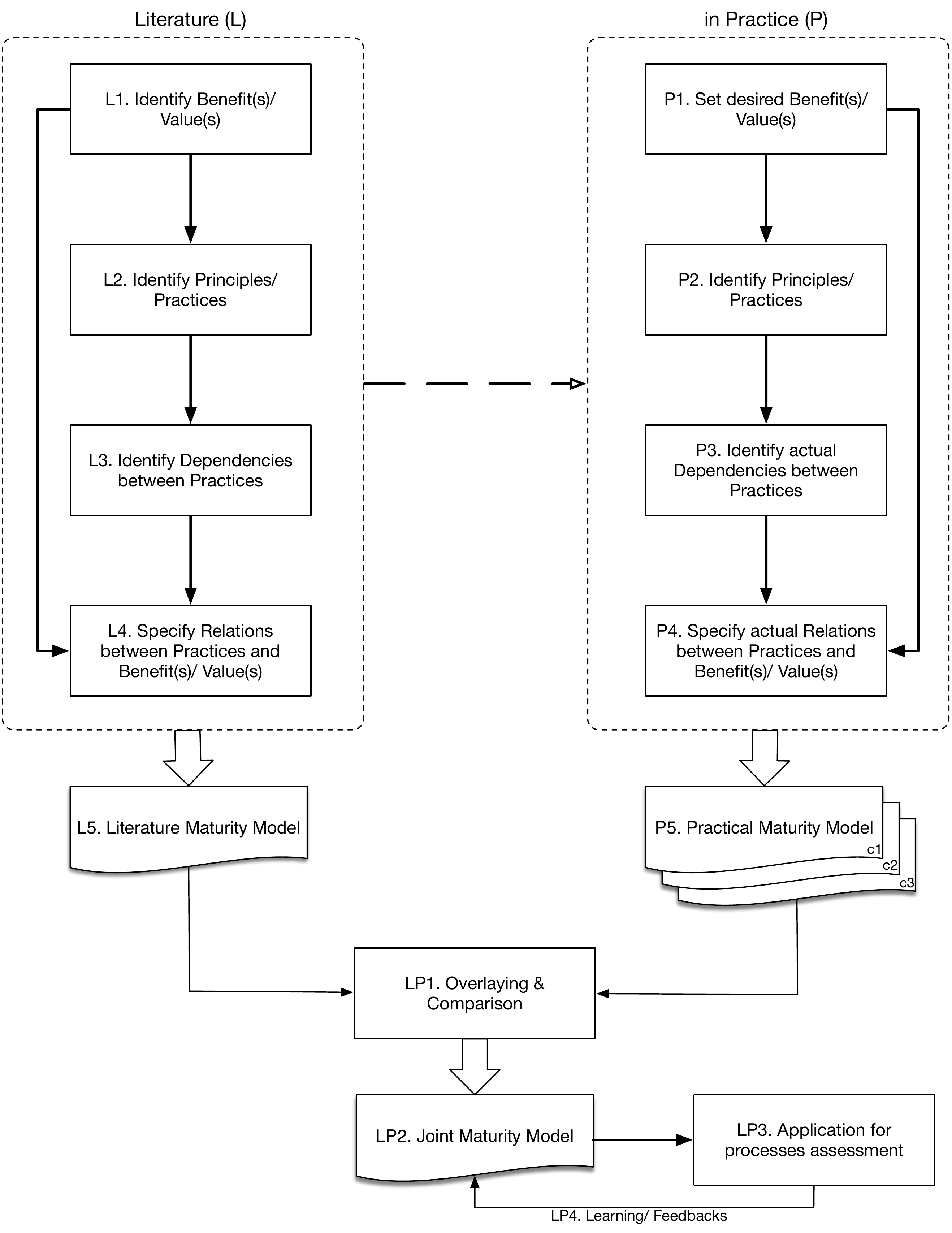}
\caption{a Value-driven Approach to design a new maturity model, VASPI}
\label{fig:guidelineoverview}
\end{figure}

\subsubsection{Literature approach (L)}
First, we follow the L approach in order to design a maturity model (L5) based on findings would be extracted from literature (peer-reviewed literature, gray literature, or technical reports). To create L5, the following steps have been taken:
\begin{itemize}
\itemsep1em
\item \textbf{L1. Identify Value(s)/Benefit(s):} According to the context in which a new model would be designed, the desired values and benefits have to be identified from relevant literature.

For example, to create a maturity model for a specific agile context \textit{e.g., software deployment}, the desired values and benefits of \textit{deployment context} have to be identified.

Several methods exist to conduct a literature study in accordance with needs and concerns. For example, in software engineering, the most well-known methods for performing literature reviews are Systematic Literature Reviews (that could be conducted based on the guidelines proposed by Kitchenham \textit{et al.} \cite{SLRguidelines}) and Systematic Mapping Study (two samples of the guideline are \cite{SLRguidelines} by Kitchenham \textit{et al.} and \cite{Petersen2015guidelines} by Petersen \textit{et al.}). Garousi \textit{et al.} \cite{multivocalreview} also present a method, Multivocal Literature Review, that includes gray literature as well as peer-reviewed literature.

Primary studies (original research), secondary (systematic reviews) and tertiary literature (handbooks) \cite{SLRguidelines}, peer-reviewed literature (scientific/academic peer-reviewed journals, and peer-reviewed conferences articles), and nonpeer-reviewed ones like technical reports, and blogs could be used as sources.

As an example, by the use of systematic literature review \cite{SLRguidelines} and conducting a study on `deployment', the following benefits related to the \textit{deployment} might be identified: \textit{Cost saving, Fast and frequent release, Quick responses, Increase productivity, Predictability, and Repeatability} \cite{76,111,39,4}

Then, the identified benefits can be linked to the corresponding software values to provide a common understanding of different value perspectives (e.g., customer, business). To this end, Software Value Map (SVM) \cite{khurumsoftwarevaluemap} presented by Khurum \textit{et al.} can be used. SVM provides an exhaustive view on value perspectives (Financial, Customer, Internal Business Process, Innovation and Learning) as a unified collection that indicates a common understanding to ensure not missing any perspective \cite{khurumsoftwarevaluemap}. Table \ref{tab:svmexample} shows an example of SVM and a value perspective and its value aspects, sub aspects and components. Also, Table \ref{tab:examplebenefitsvalues} shows an example of the identified benefits categorized by the use of SVM. 

\begin{table}[!ht]
\centering
\caption{An example of SVM \cite{khurumsoftwarevaluemap}}
\label{tab:svmexample}
\scalebox{0.9}{
\begin{tabular}{c p{3.6cm}p{3.4cm}p{3.1cm}}
\textbf{Value Perspective} & \textbf{Value Aspect} & \textbf{Sub-value Aspect} & \textbf{Value Components} \\ \hline
\multirow{5}{*}{Customer} & \multirow{4}{*}{Perceived value} & \multirow{3}{*}{Intrinsic value} & functionality \\ \cline{4-4} 
 &  &  & reliability \\ \cline{4-4} 
 &  &  & usability \\ \cline{3-4} 
 &  & Delivery process value & process w.r.t. time \\ \cline{2-4}
 & Customer lifetime value & Revenue & upselling revenue \\ \hline
\end{tabular}}
\end{table}

\begin{table}[!ht]
\centering
\caption{An example of benefits categorized by SVM}
\label{tab:examplebenefitsvalues}
\scalebox{0.9}{
\begin{tabular}{p{4.2cm}p{6.5cm}p{4cm}}
\multicolumn{1}{c}{\textbf{Benefits (Bn)}} & \multicolumn{1}{c}{\textbf{Value perspective}} & \multicolumn{1}{c}{\textbf{Value component}} \\ \hline
\begin{tabular}[c]{@{}l@{}}B1. Fast \& frequent releases\\ B2. Cost saving\\ B3. Quick responses\end{tabular} & Customer/ perceived value & \begin{tabular}[c]{@{}l@{}}$\bullet$ functionality\\ $\bullet$ reliability\\ $\bullet$ usability\end{tabular} \\ \hline
B4. Increase productivity & Internal Business Process/ production value & \begin{tabular}[c]{@{}l@{}}$\bullet$ market requirement value\\ $\bullet$ physical value wrt. time\end{tabular} \\ \hline
\begin{tabular}[c]{@{}l@{}}B5. Repeatability\\ B6. Predictiability\end{tabular} & Innovation and learning/ value of technology & \begin{tabular}[c]{@{}l@{}}$\bullet$ human capital value\\ $\bullet$ customer capital value\\ $\bullet$ market value size\end{tabular} \\ \hline
\end{tabular}}
\end{table}

\item \textbf{L2. Identify Principles/Practices:} This step attempts to identify principles which exist to support the values of the context specified in L1. Also, corresponding practices that promise the specified benefits would be identified in L2.

The importance of identifying principles is to characterize the context and knowing the fundamentals. Also, it assists to better understand values and recognize practices, in which the principles have been applied \cite{principles}. Identifying the principles helps us to find relevant practices and realize alternative practice sets.

\item \textbf{L3. Identify Dependencies between Practices:} 
A process of software development contains a number of applied practices. Consequently, to identify the actual practices that need to be in place to achieve the desired benefits, the relations among practices and the alternatives should be identified.

For example, a number of practices can be identified such as \textit{continuous deployment} \cite{111}, \textit{automated deployment}, and \textit{continuous integration} \cite{48} for \textit{deployment}. But, how these practices should be in place is missing. Figure \ref{fig:practicesrelations} shows one alternative formation for the practices according to the identified relations from the literature (R1 \cite{111} and R2 \cite{48}).

\begin{figure}[ht]
\centering
\includegraphics[scale=0.4]{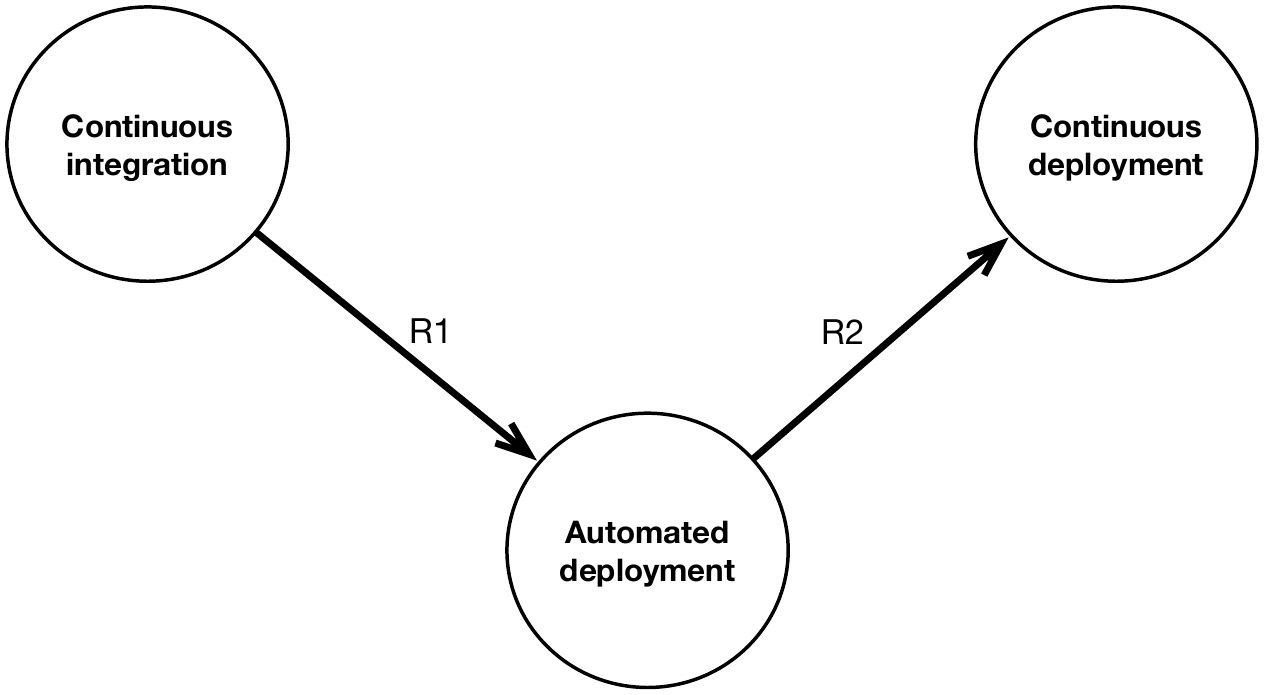}
\caption{An example of identifying the relations between specified practices}
\label{fig:practicesrelations}
\end{figure}

\item \textbf{L4. Specify Relations between Practices and Value(s)/Benefit(s):}
Before step L4, values/benefits (L1), practices (L2), and relations among practices (L3) are specified. Step L4 aims to realize the link of practices to the desired values/benefits through the same procedure performed in L3 (extracting relevant data from the literature).

By specifying the desired benefit(s) of a practice or a set of practices, it is possible to trace the link from practice(s) to benefit(s) and vice versa to pursue the desired goals and values.

One initial strategy to specify relations between practices and values/benefits can be (1) identifying the expected benefit(s) of each practice and then, (2) aggregating the identified benefits as an outcome from a set of practices. Figure \ref{fig:practicesbenefits} shows an example of relations between practices and benefits. 

\begin{figure}[ht]
\centering
\includegraphics[scale=0.45]{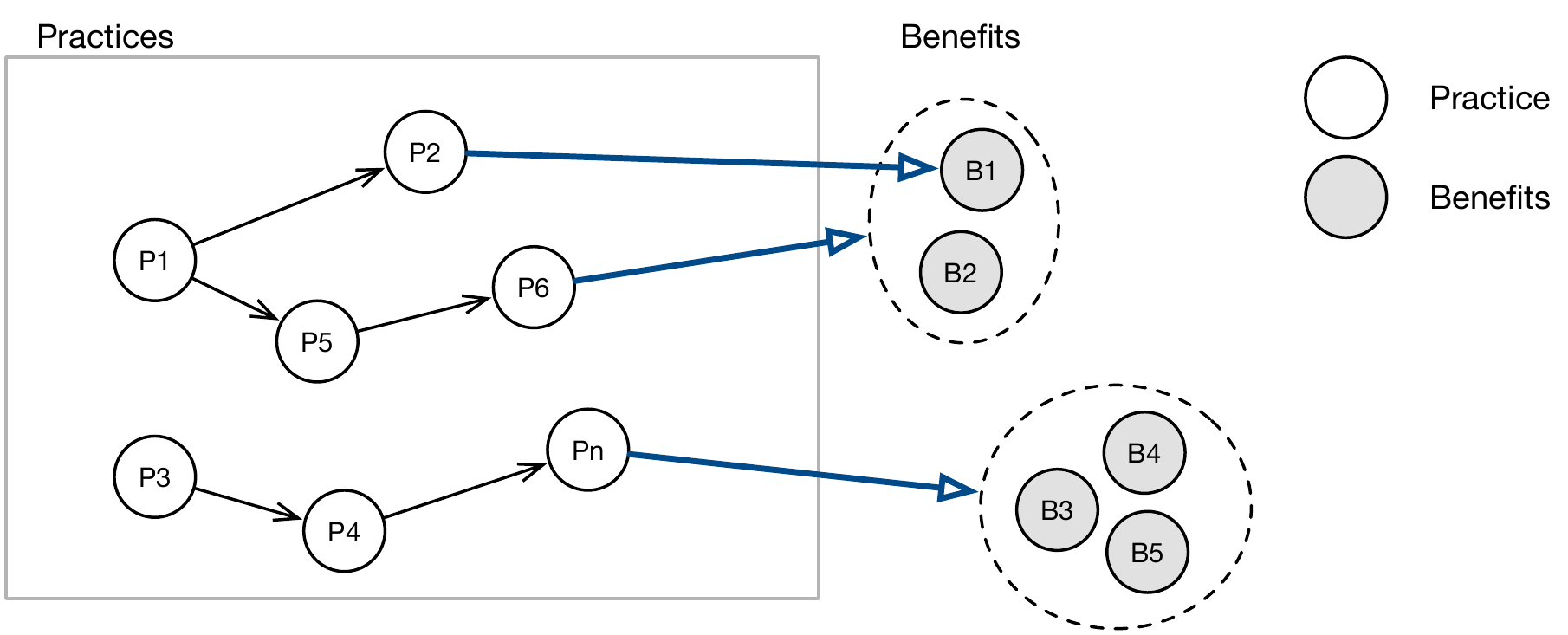}
\caption{An example of relations between practices and benefits}
\label{fig:practicesbenefits}
\end{figure}
\end{itemize}

For example, Fast and frequent releases (B1), Quick responses (B3), Increase productivity (B4), Predictability (B6), and Communication/Collaboration have been identified as the desired benefits of the practice \textit{Continuous integration} \cite{111,39,4}.

\begin{figure}[H]
\begin{adjustbox}{addcode={\begin{minipage}{\width}}{\caption{%
      An example of a maturity model derived from research literature (L5)
      }\label{fig:exampleLMM}\end{minipage}},rotate=90,center}
\includegraphics[height=0.4\textheight]{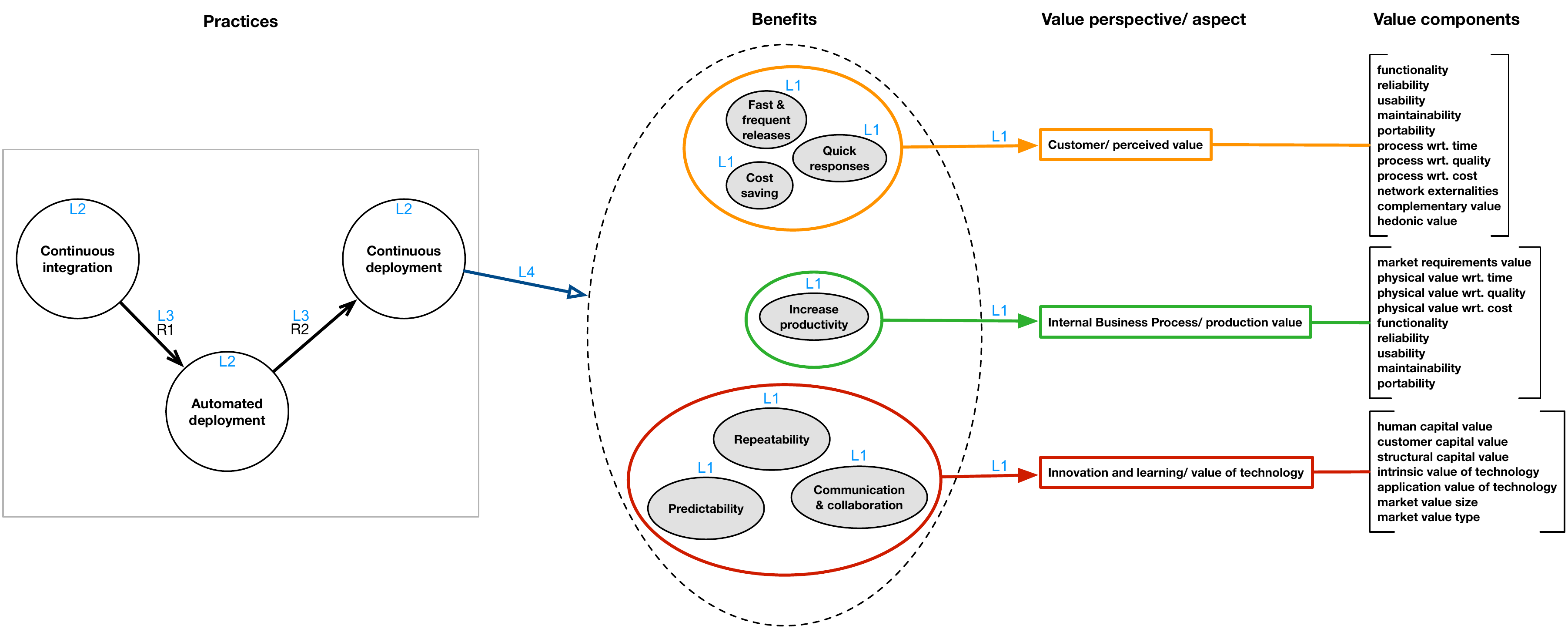}
\end{adjustbox}
\end{figure}

To provide a consolidated view of the whole approach and in order to create an instance of Maturity Model derived from literature (L5), the complete findings including \textit{values/benefits}, \textit{practices and dependencies}, and \textit{relations between practices and benefits} have to be formed together. Figure \ref{fig:exampleLMM} shows a sample of L5 created by the guideline of Literature approach in this study. It should be mentioned that this is not an order in which the practices are executed but rather the dependencies. A dependency indicates how practices have to be in place to achieve the desired benefits/values.

\subsubsection{in-Practice approach (P)}
We follow the in-Practice approach to design a maturity model (P5) based on findings would be from empirical studies and methods such as workshops, focused groups, and case studies. To create P5 the following steps have to be taken:

\begin{itemize}
\itemsep1em
\item \textbf{P1. Set desired Value(s)/Benefit(s):} According to the context in which a new model would be designed, the desired values and benefits have to be set. The desired values and benefits can come from business plans, organizational goals, running workshops with stakeholders, etc.

\item \textbf{P2. Identify Principles/Practices:}
This step attempts to identify principles which exist to support the values of the context specified in P1. Also, corresponding practices that promise the specified benefits would be identified in P2.

P2 supports what has been identified in L2 and indicate what is missing. Gap analysis, interviews, observations, workshops can be used as a means for data collection.

\item \textbf{P3. Identify actual Dependencies between Practices:}
After specifying the practices, it is turn to position them based on their relations or dependencies. It helps an organization, project, or product team to concentrate on real concerns and actual activities according to the desired goals without any disadvantage of traditional maturity models such as overheads of cost, time, effort, additional documentations, and resources \cite{disadvantageCMMI}.

\item \textbf{P4. Specify actual Relations between Practices and Value(s)/Benefit(s):}
Before P4, values/benefits (P1), practices (P2), and relations among practices (P3) have been specified. Step P4 aims to realize the link of practices to the desired values/benefits to create a practical maturity model (P5) in accordance with actual case(s) (c1, c2, and c3 in Figure \ref{fig:guidelineoverview}).

By specifying the desired benefit(s) of a specific practice or a set of practices, it would be possible to trace the link from practice(s) to benefit(s) and vice versa to pursue the desired goals and values.

\end{itemize}

\subsubsection{Joint Maturity Model}
By mapping relevant instances of L5 and P5 and making comparison between similarities and differences (LP1) of the context, concerns, values, practices, and relations, it is possible to create a joint sample of maturity model.

LP2 aims to bring the state of the art and practice together to provide a holistic view on development processes. Then, the joint model can be used in process assessment activities in accordance with real concerns (LP3). Based on the application of a joint model in a specific context, a corresponding joint maturity model would be incrementally evolved through learning and feedback (LP4).

\section{Conclusion and Future Research}
\label{sec:conclusionandfuture}

In this paper, we presented VASPI as a value-driven approach for software process improvement. Through a literature review, we identified existing approaches for maturity and process assessments. We have analyzed these existing models, their purpose, structure, and guidelines. Moreover, we considered the relations and dependencies between practices and their links to the desired benefits. To this end, four research questions have been answered. 

\vspace{5mm}

\textit{RQ1. What is the generic structure of maturity models presented in the literature?}

To address the RQ1, we reviewed the structure of maturity models presented in the literature. We identified their type and specification such as levels of maturity, descriptors, dimensions, and activities. However, the majority of discussions have been about staged models like Crosby, CMM, and CMMI \cite{maturityclassification,crosby,maturitygrids,CMM,lasrado2015maturity,13criticalreview}. The common characteristics (Number of levels, Descriptor of levels, Descriptions, Dimensions, and Activities) of the maturity models have also been identified. In addition, we looked at the structure of CMMI as the most used staged maturity model. The staged models most often suffer from a lack of ability for expressing dependencies among processes within a capability level, while flexible models allow the definition of intermediate status that indicates further details of maturity \cite{devmmslr,29spiforvse,27focusareaMM}.

\vspace{5mm}

\textit{RQ2. What are the approaches/guidelines/methods to design a new maturity model introduced in the literature?}

To address the RQ2, we found studies which propose new maturity models. Through our investigation, we realized that new models most often suffer from avoiding existing solutions, a lack of theoretical basis, empirical studies, and validation \cite{devmmslr,maturitymodelmappingstudy,lasrado2015maturity,13criticalreview}. We also identified the methods presented in the literature to design a new maturity model. Although the identified methods have not been presented in detail or they belong to a specific context, we have specified the main aspects of the procedure for designing new models. The generic activities performed to create a new maturity model have been aggregated through considering six central characteristics (Context, Origination, Source, Type, Meta-model, and Evaluation).

\vspace{5mm}

\textit{RQ3. To what degree do existing maturity models consider value/ benefit into account for assessment?}

The majority of existing models have been limited to indirect measurement and evidence-based models like CMMI and formal models. They do not consider the relations between practices and desired benefits and quality attributes \cite{22valuematurity}. These models focus on software quality attributes in general (like efficiency, effectiveness, and suitability) which are not linked to every single practice within the software process development. Some maturity models like CollabMM consider quality attributes and benefits as advantages of using the model but do not specify the relations between practices and benefits. \cite{24CollabMM}. We also identified the OPP framework (Objectives, Principles, and Practices) that considers the relation between principles and practices but it does not define the relations among practices \cite{soundararajan2011}.

We conclude that there is a lack of models to assess a set of practices by their desired benefits. The majority of existing models such as CMMI, SPICE, RPI, BPMM, and KMMM (see Table \ref{tab:comparisonsteps}) are fixed model type with rigid levels of maturity that they cannot define interdependencies among different types of practices. In addition, they are costly to implement \cite{25baars2016,9spiSLR,29spiforvse,27focusareaMM}.

\vspace{5mm}

\textit{RQ4. How could a new maturity model be proposed for a value-based assessment approach?}

The concept of maturity models has been mostly concluded in traditional approaches like CMMI that assess maturity based on predefined standards and frameworks. However, the movement towards being agile points to a need for maturity models that fit agile practices. Although a number of models have been presented in this context, they suffer from a lack of holistic view of the whole process. Furthermore, these models do not explicitly consider values, benefits, practices, and relations among them.

To address the gaps identified in this study, we realized the need for a new approach for the maturity that enables us to meet a lack of actual relations between practices and desired values/benefits with a consolidated view on the whole software development process (See section \ref{subsubsec:newmaturitymodel}). We have proposed a value-driven approach for software process Improvement (VASPI) through conducting literature reviews to design a new maturity model in accordance with real concerns. By the use of VASPI, it would be possible to look at practices, desired benefits, and corresponding values at the same time as it attempts to identify relations among them in particular context.

\vspace{5mm}

In the future, we will apply `Literature approach' of VASPI for DevOps by including gray and peer-reviewed literature (DevOps-VASPI). We also intend to apply the `In-practice approach' in cooperation with a software company. We will use empirical research methods to evaluate and validate VASPI.

\section*{Acknowledgment}
This work has been supported by ELLIIT, a Strategic Area within IT and Mobile Communications, funded by the Swedish Government. The work has also been supported by research grant for the ViTS project (reference number 20180127) from the Knowledge Foundation in Sweden.

\bibliographystyle{splncs}
\bibliography{references.bib}


\end{document}